\documentclass{iopconfser}
\usepackage{graphicx}
\usepackage{siunitx}
\usepackage{amsmath}
\usepackage{subfigure}
\usepackage{booktabs}

\begin{document}

\title{Measurements of dislocations in 4H-SiC with rocking curve imaging}

\author{Ahmar Khaliq$^{1,2}$, Felix Wittwer$^{1,2}$, Niklas Pyrlik$^{2}$, Giovanni Fevola$^{2}$, Svenja Patjens$^{2}$, Jackson Barp$^{2}$, Gero Falkenberg$^{2}$, Sven Hampel$^{2}$, Michael Stuckelberger$^{2}$, Jan Garrevoet$^{2}$, Dennis Brückner$^{2}$, Peter Modregger$^{1,2*}$}

\affil{$^{1}$Physics Department, University of Siegen, Walter-Flex-Str. 3, Siegen, 57072, Germany}
\affil{$^{2}$Center for X-ray and Nano Science, Deutsches Elektronen-Synchrotron, Hamburg, 22607, Germany}

\email{ahmar.khaliq@desy.de}

\begin{abstract} 
4H Silicon Carbide (4H-SiC) combines many attractive properties such as a high carrier mobility, a wide bandgap, and a high thermal conductivity, making it an ideal candidate for high-power electronic devices. However, a primary challenge in utilizing 4H-SiC is the presence of defects in epitaxial layers, which can significantly degrade device performance. In this study, we have used X-ray transmission topography with a rocking curve imaging technique to characterize the types and distribution of defects in 4H-SiC. The derived maps from the fitted Gaussian parameters were used to investigate dislocations in 4H-SiC. Understanding the distribution of the dislocations provides valuable insights into the overall crystal quality, which can guide improvements for the fabrication processes.
\end{abstract}

\section{Introduction}

Silicon carbide (4H-SiC) is a promising semiconductor material owing to its wide band gap, high thermal conductivity, and remarkable chemical stability \cite{li2022dislocations}. These characteristics make 4H-SiC attaractive for high-power electronic devices \cite{kimoto2014fundamentals}. However, a primary challenge in utilizing 4H-SiC is the presence of defects in epitaxial layers, which can significantly degrade the device performance \cite{berechman2010electrical, skowronski2006degradation}. These defects can be categorized as either growth-induced, occurring during the crystal growth process, or stress-induced, arising from external stresses during wafer processing and device operation \cite{chen2008defects}.

\begin{figure}
    \centering
    \includegraphics[scale=0.17]{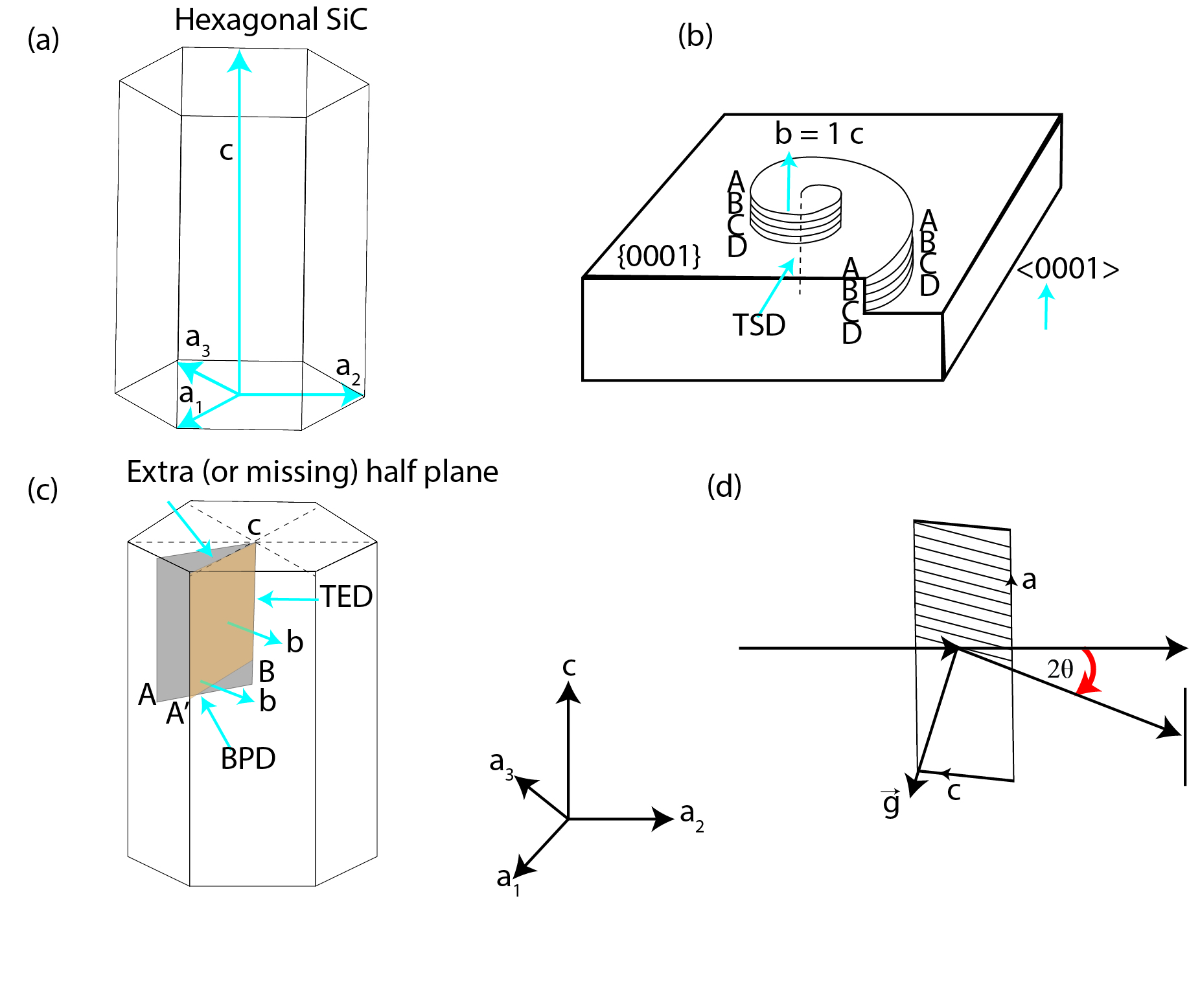}
    \caption{(a) Schematic drawing of a hexagonal unit cell and primary translation vectors of 4H-SiC along the axial and basal directions are defined as \textbf{c} and \textbf{a}. 
    (b) An elementary TSD in 4H-SiC is located at the center of a spiral with a four-bilayer-height step (ABCD) corresponding to $|1\textbf{c}|$. (c) Here an extra (or missing) half-plane with two types of dislocations in a 4H-SiC crystal. The dislocation along horizontal line AB (basal plane) is a BPD along the ⟨11$\Bar{2}$0⟩ direction, while that along vertical line BC ([0001] direction) is a TED. (d) Typical configuration of an X-ray diffraction setup, illustrating the incident beam, transmitted diffracted beam, and the diffraction vector \(\mathbf{g}\). \cite{kimoto2014fundamentals}.} 
    \label{defects}
\end{figure}

A common type of defects in 4H-SiC epitaxial layers are dislocations. Dislocations can be grouped as threading screw dislocations (TSDs), threading edge dislocations (TEDs), and basal plane dislocations (BPDs) \cite{li2022dislocations}. Figure \ref{defects} illustrates the basic structures of the 4H-SiC lattice and dislocations \cite{kimoto2014fundamentals}. It has been established that these dislocations are typically generated during the growth phase of 4H-SiC single crystals \cite{shinagawa2020populations, hoshino2017fast}. The nature of dislocations is determined by the Burgers vectors \(\mathbf{b}\) and dislocation line directions \(\mathbf{l}\) \cite{kimoto2014fundamentals}, as summarized in Table \ref{vectors}. The visibility of dislocations depends upon diffraction condition \( \mathbf{g} \cdot \mathbf{b} \neq 0 \), where \(\mathbf{g}\) is the the diffraction vector \cite{zhang2024synchrotron, raghothamachar2010x}. These dislocations lead to a reduction in the lifetime of minority carriers, promote leakage currents, and cause reliability issues, thereby affecting the performance of 4H-SiC-based devices \cite{kimoto2020defect}.





To improve the performance and reliability of these devices, it is essential  to identify dislocation types \cite{kamata2018x} 
and evaluate their impact on electrical performance \cite{fujiwara2012relationship, wahab2000influence}. Typical methods to investigate dislocations in 4H-SiC epilayers include potassium hydroxide (KOH) etching \cite{katsuno1999mechanism}, transmission electron microscopy (TEM) \cite{sato2014dislocation, twigg2004partial}, scanning electron microscopy \cite{sato2014dislocation}, and electron beam-induced current \cite{yao2011simultaneous}. KOH etching reveals the locations and types of dislocations, but it is challenging to characterize the directions of the Burgers vectors of these dislocations. Similarly, TEM can provide valuable information on the ultrastructure, morphology, and the nature of the bonds of the sample. However, it is a destructive technique that can not be used for in-situ or real-time characterization. 

\begin{table}
\centering
\caption{Overview of Burgers vectors, dislocation line orientations, and dislocation densities in 4H-SiC single crystals \cite{kimoto2014fundamentals}.}
\begin{tabular}{lccc}
\toprule
\textbf{dislocations} & \textbf{burgers vectors} & \textbf{direction of dislocation-line} & \textbf{typical density (cm$^{-2}$)} \\
\midrule
TSDs & \textbf{c} & [0001] & $10^{2}$ - $10^{3}$ \\
TEDs & [11$\Bar{2}$0]/3 & roughly parallel with [0001] & $10^{3}$ - $10^{4}$ \\
BPDs & [11$\Bar{2}$0]/3 & (0001) plane (predominantly along 11$\Bar{2}$0]) & $10^{2}$ - $10^{4}$ \\
\bottomrule
\end{tabular}
\label{vectors}
\end{table}

In this work, we have performed monochromatic synchrotron radiation based X-ray topography (XRT) with rocking curve imaging (RCI) in transmission geometry to study the dislocations in a 4H-SiC crystal. 

\section{Experimental Setup}
\begin{figure}
   \centering
   \includegraphics[scale=0.25]{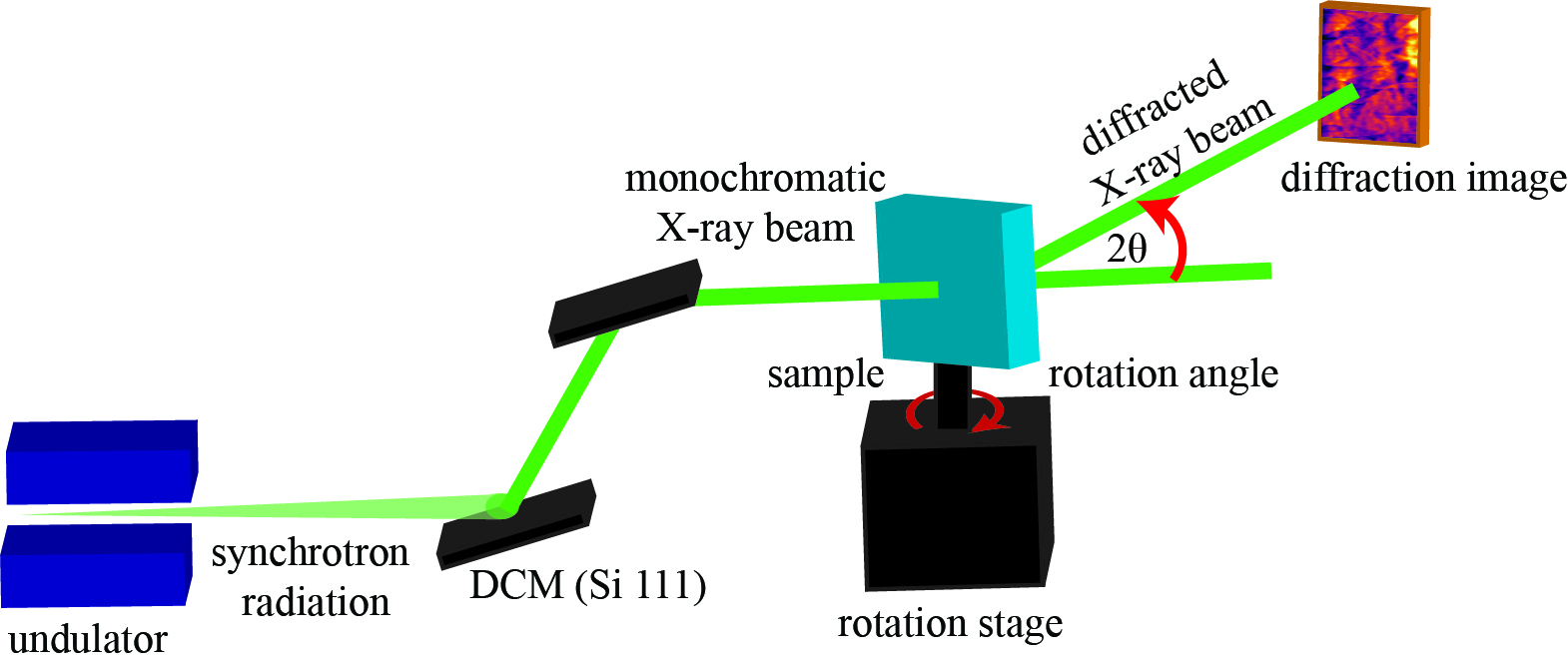}
   \caption{Sketch of the experimental setup.}
   \label{setup}
\end{figure}

The Beamline P06 at PETRA III, DESY, Hamburg, Germany was used for this study \cite{schroer2010hard}. In Figure \ref{setup}, a sketch of the experimental setup is shown. An X-ray beam of 20 keV was selected using a Si(111) double-crystal monochromator (DCM). The beam was adjusted to size of $\SI{500}{\micro\meter} \times \SI{500}{\micro\meter}$ and aimed onto a SiC sample placed on a high-precision six-axis goniometer (SmarAct) \cite{chakrabarti2022x} with nominal angular resolutions below $\SI{0.1}{\micro\radian}$. The x, y, and z translation stages on top of the rotation axes enabled a nominal position accuracy of $\SI{1}{\nano\meter}$ for horizontal alignment along the beam (x) and scanning of the sample (y and z) with fixed rotation angles. The RCI technique was used to perform the measurements. The samples were exposed to the incident beam for 2 seconds in transmission geometry in the (100) direction, where the sample was rocked in a range of -0.1 to 0.1 degrees (step size: $\SI{0.001}{\degree}$, total number of steps: 201). Diffraction topograms were recorded with an X-ray camera (optique peter microscope + PCO edge detector). The detector had an effective pixel size of $ \SI{0.5}{\micro\meter}$ and a field of view of $ 1 \times \SI{1}{\mm}$.


\section{Data Analysis}

\begin{figure}
\centering
\includegraphics[scale=0.18]{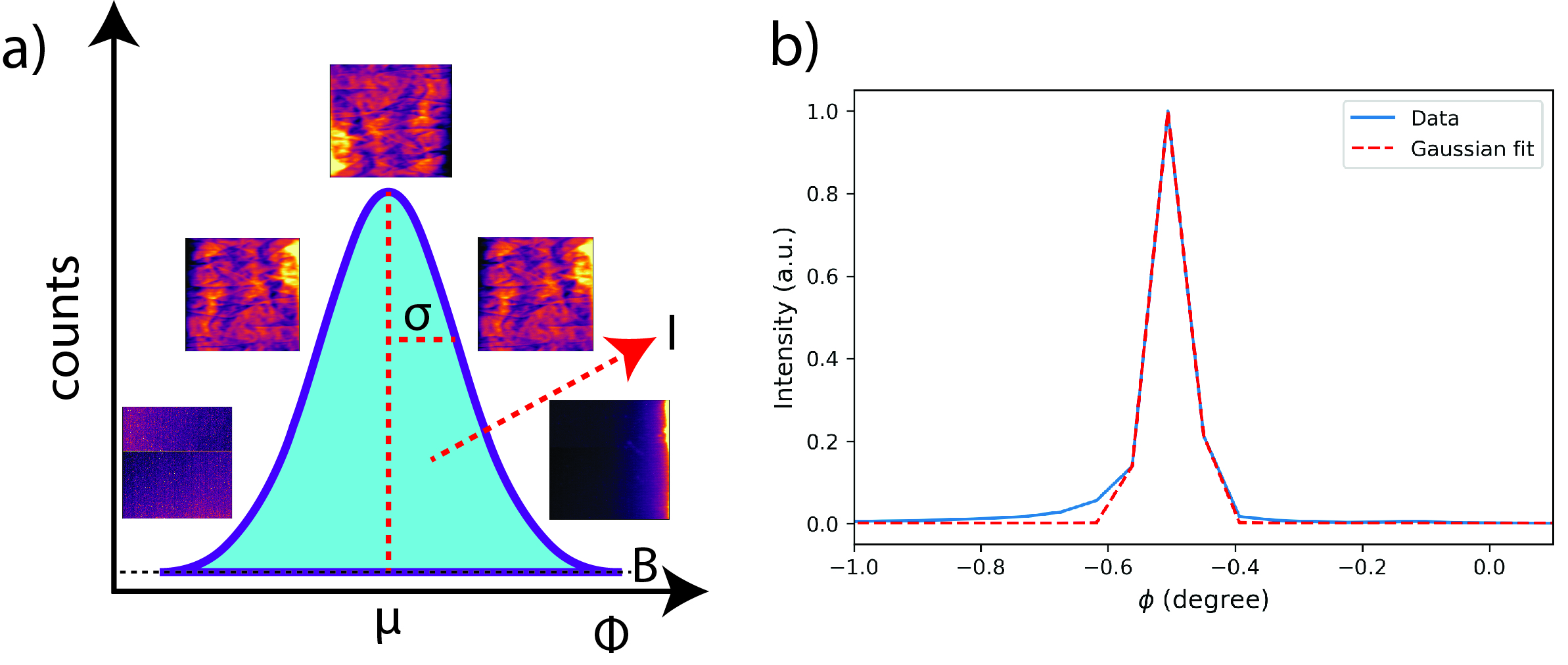}
\caption{(a) Acquisition of an image series, each pixel records its own local rocking curve. A schematic drawing of the fitting procedure of integrated intensity \( I \), peak position \(\mu\), and peak width \(\sigma\), extracted on a pixel-by-pixel basis from the rocking curves. (b) Fitting example for one pixel.}
\label{fig:acquisition_series}
\end{figure}

The RCI technique involves capturing multiple images while rotating the sample in angular steps through the Bragg diffraction angle, as illustrated in Figure~\ref{fig:acquisition_series}(a). An image is recorded at each step, resulting in a stack of images where each pixel contains a local rocking curve. 


\begin{equation}
f(\theta) = A \exp \left( -\frac{(\theta - \mu)^2}{2\sigma^2} \right) + B
\label{eq:gaussian}
\end{equation}

where:
\[
I = \int_{-\infty}^{\infty} f(\theta) - B \, d\theta
\]

Each local rocking curve was independently fitted using a Gaussian function described in (1), where \( I \) corresponds to the integrated intensity observed at the Bragg angle, \(\theta\) is the rocking angle, \(\mu\) is the local Bragg angle in each pixel, \(\sigma\) represents the peak width, and B is the background offset. Figure~\ref{fig:acquisition_series}(a) also shows schematic representation for evaluating the maps of the integrated intensity I, peak position $\mu$, and peak width \(\sigma\) of the diffracted image. Figure~\ref{fig:acquisition_series}(b) illustrates the comparison of the fitting process for individual pixels, where the fitted Gaussian curve closely matches the experimental data points. 


\section{Discussion}

\begin{figure}
    \centering
    \begin{minipage}[b]{0.323\textwidth}
        \centering
        \includegraphics[width=\textwidth]{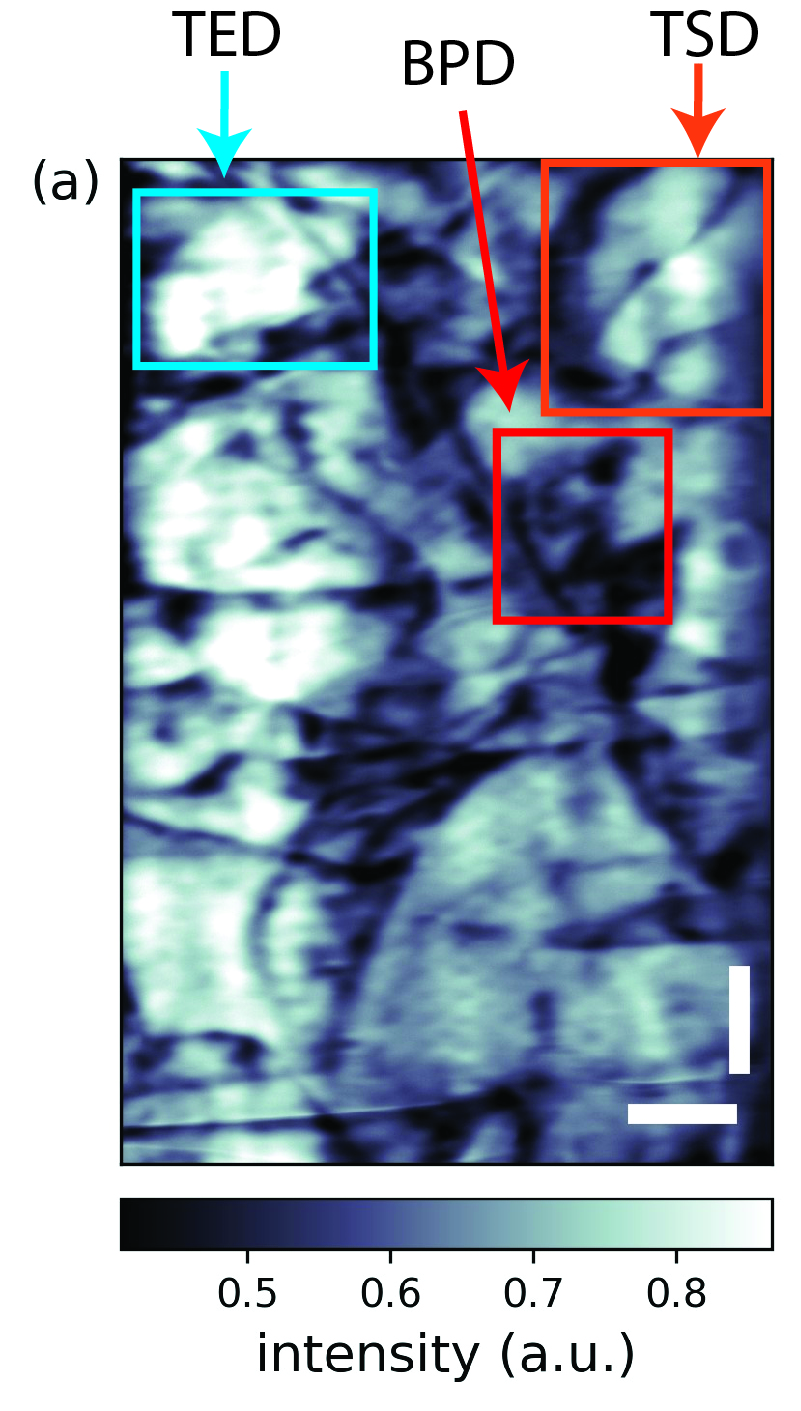}
    \end{minipage}
    \hspace{-0.02\textwidth} 
    \begin{minipage}[b]{0.326\textwidth}
        \centering
        \includegraphics[width=\textwidth]{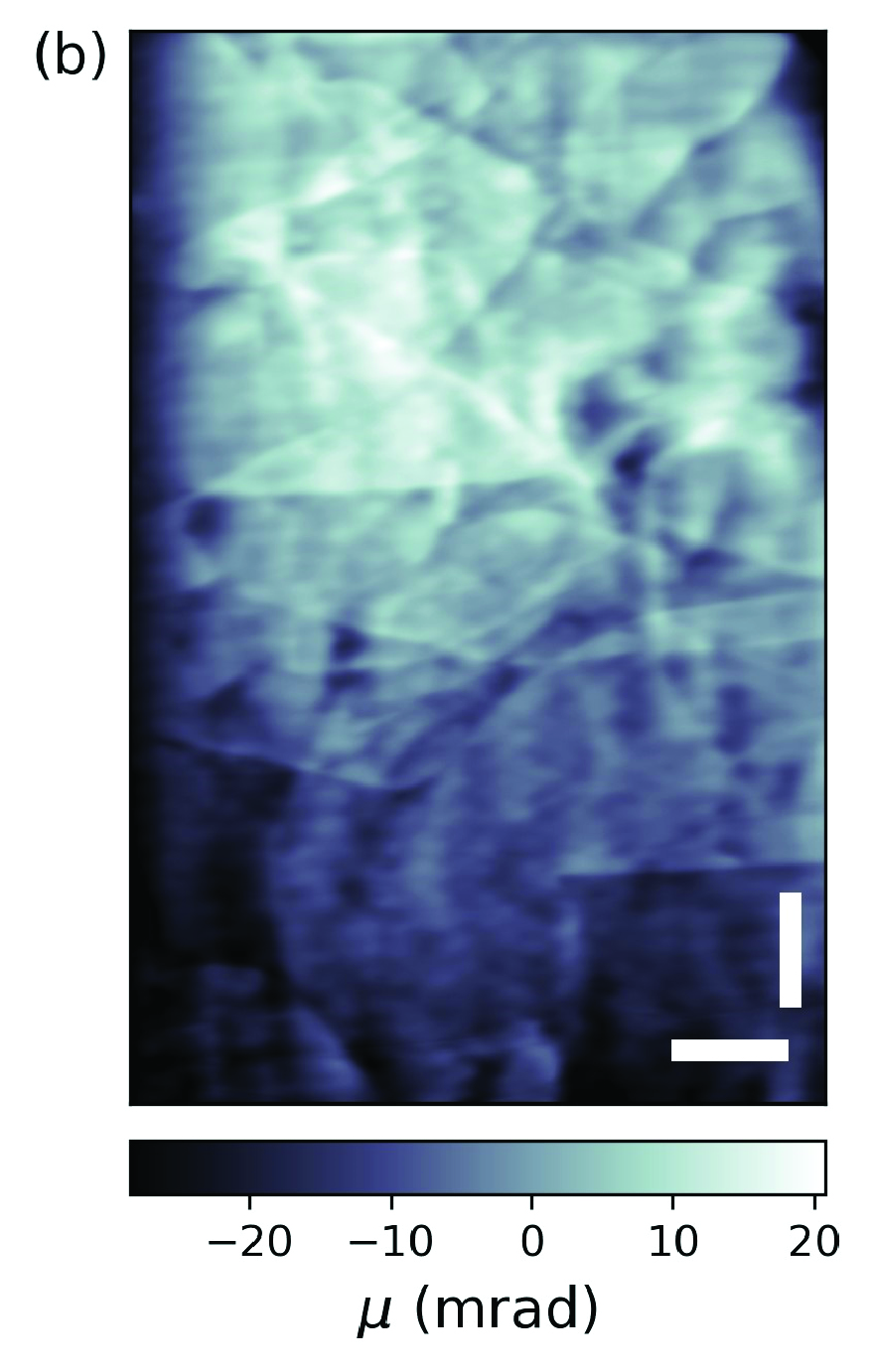}
    \end{minipage}
    \hspace{-0.02\textwidth} 
    \begin{minipage}[b]{0.33\textwidth}
        \centering
        \includegraphics[width=\textwidth]{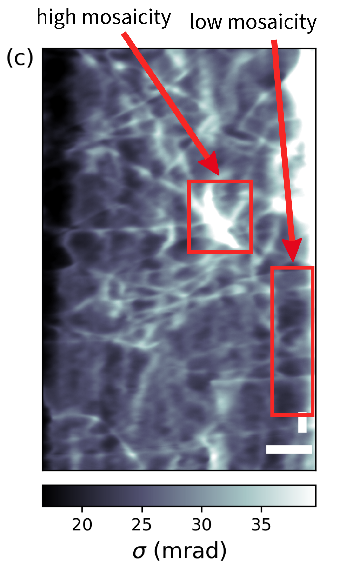}
    \end{minipage}
    \caption{(a) Resulting map of integrated intensity \( I \) showing dislocation structures. (b) Peak position map indicating gradient in \( \mu\). (c) \( \sigma\) map reflecting the lattice quality. The horizontal bar represents $\SI{400}{\micro\meter}$, and the vertical bar represents $\SI{800}{\micro\meter}$.}
    \label{resultant_maps}
\end{figure}

Each type of dislocation exhibits a characteristic fingerprint in the XRT images, arising from the differences in the Burgers vectors and line directions, see Table \ref{vectors}. For example, TSDs align along the vertical plane (\textbf{c}-axis), BPDs align along the horizontal plane, i.e., perpendicular to the sample’s surface, and TEDs extend vertically along the c-axis but with different Burgers vectors compared to TSDs. The resulting maps in Figure~\ref{resultant_maps} shows 
the integrated intensity map, the peak position map, and the peak width map. The intensity map provides information on the types and distribution of dislocations in the crystal structure with circular patterns indicating TSDs, line features corresponding to BPDs, and TEDs also identifiable. The visibility of these dislocations is supported by their \( \mathbf{g} \cdot \mathbf{b} \) analysis, confirming that TSDs are visible with \( \mathbf{g} \cdot \mathbf{b} = k (1 - \cos 2\theta) \), and both TEDs and BPDs with \( \mathbf{g} \cdot \mathbf{b} = k \frac{\sin 2\theta}{6} \). The peak position map indicates variations in local Bragg angles, reflecting strain and crystal orientation \cite{authier2003dynamical}. This effective misorientation is mainly constituted by the local variation \(\Delta\phi\) in the rotation or tilt of these lattice planes and the local relative variation in the lattice parameter spacing $\Delta d/d$. The peak width map indicates the local lattice quality of the sample \cite{authier2003dynamical, tran2017synchrotron}.  

\begin{figure}
    \centering
    \begin{minipage}[b]{0.336\textwidth}
        \centering
        \includegraphics[width=\textwidth]{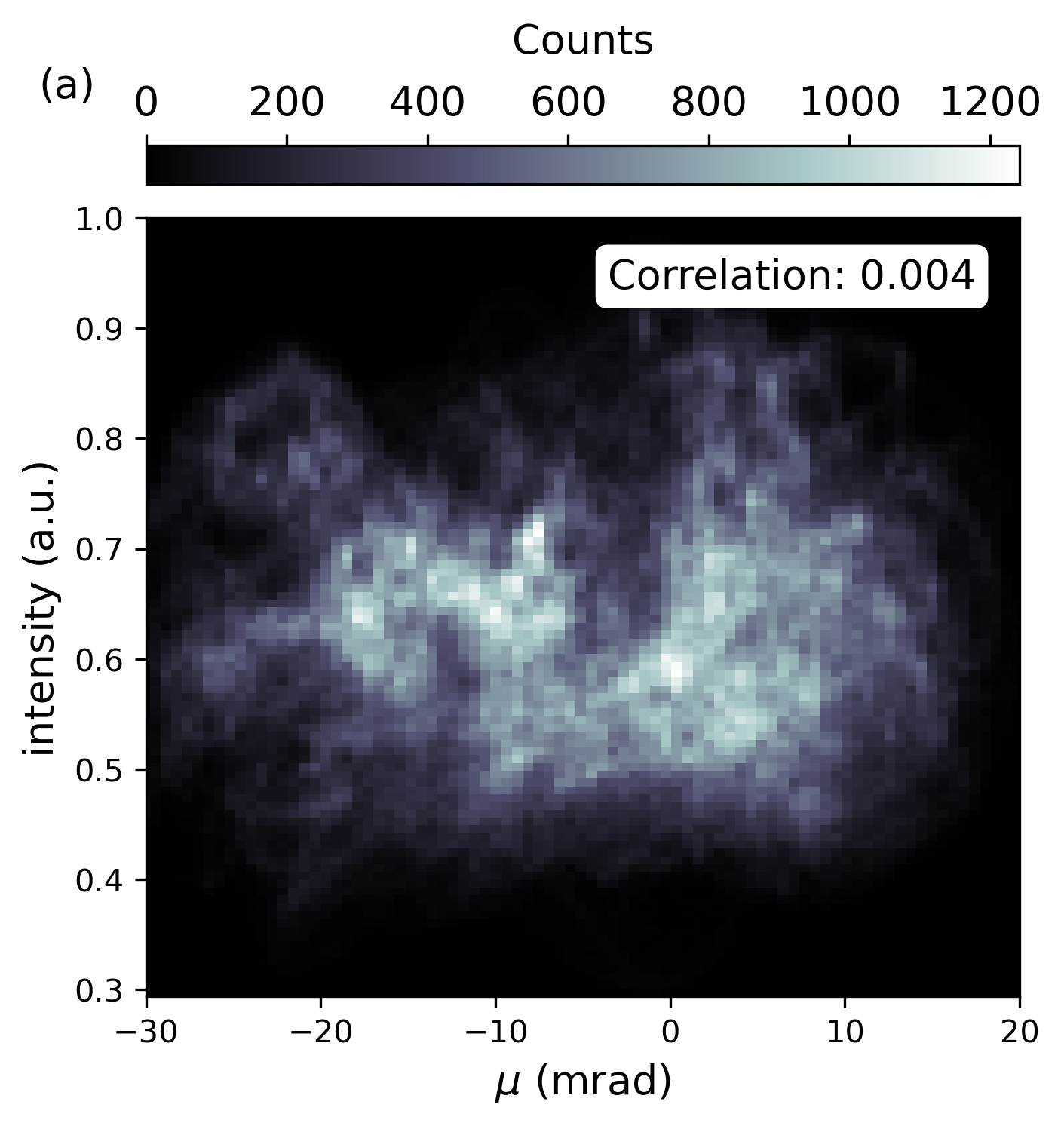}
    \end{minipage}
    \hspace{-0.02\textwidth} 
    \begin{minipage}[b]{0.33\textwidth}
        \centering
        \includegraphics[width=\textwidth]{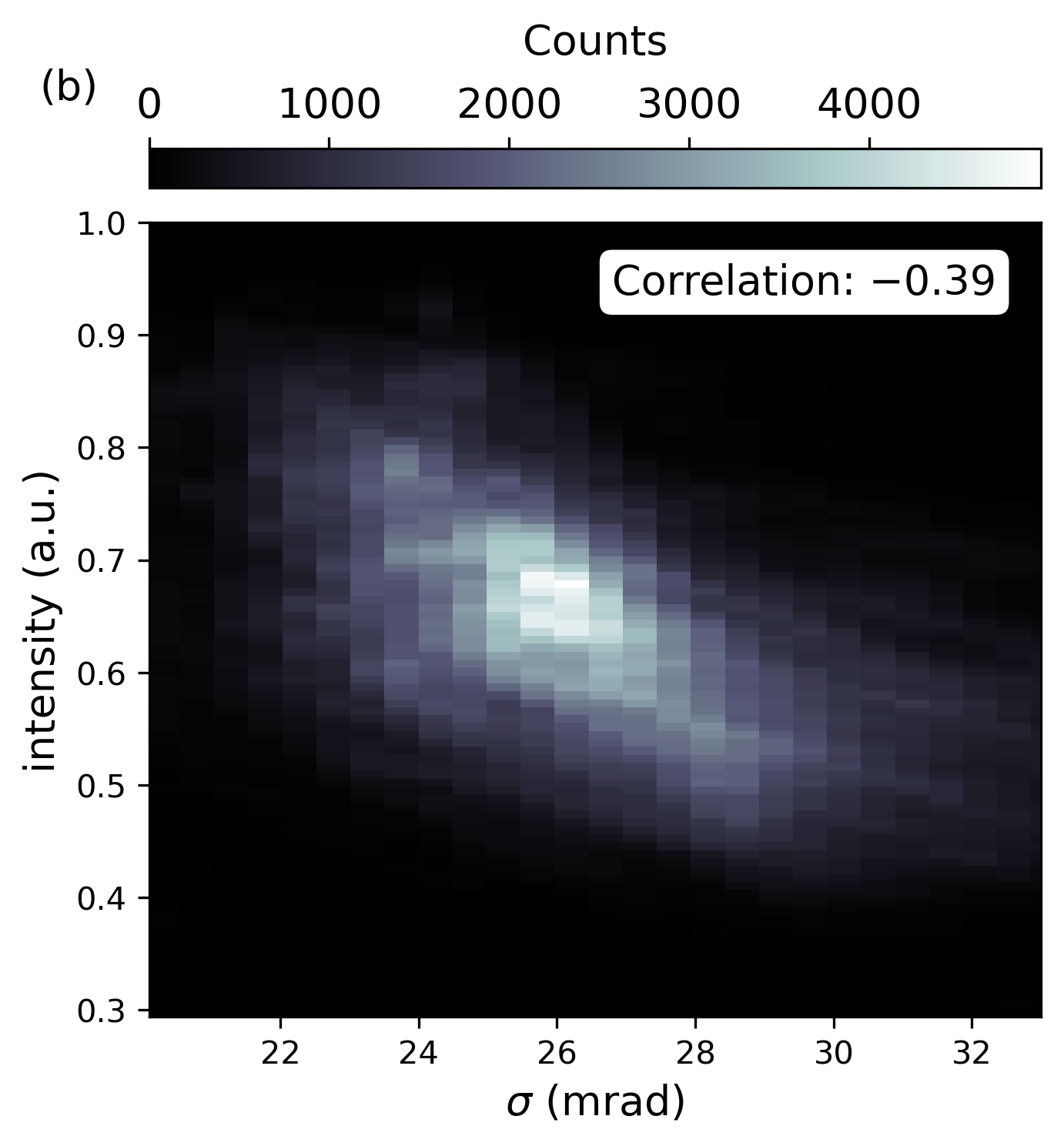}
    \end{minipage}
    \hspace{-0.02\textwidth} 
    \begin{minipage}[b]{0.3395\textwidth}
        \centering
        \includegraphics[width=\textwidth]{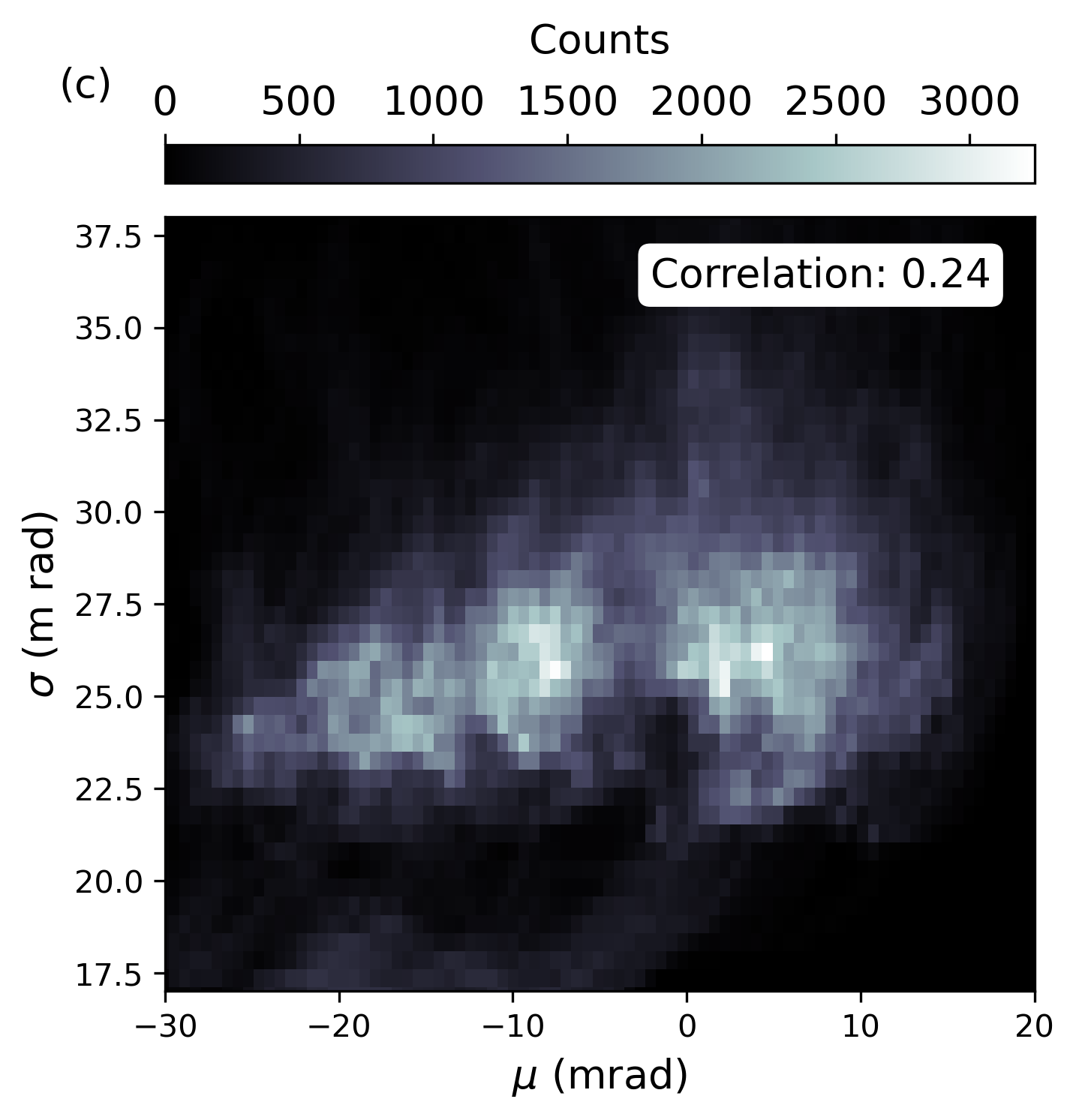}
    \end{minipage}
    \caption{The correlations between the parameters are shown: (a) correlation between integrated intensity \( I \) and \( \mu\) with a correlation value of 0.004, (b) correlation between integrated intensity \( I \) and \( \sigma\) with a correlation value of -0.39, and (c) correlation between \( \mu\) and \( \sigma\) with a correlation value of 0.24.}
    \label{correlation}
\end{figure}

The trend observed in the intensity map see Figure~\ref{resultant_maps}(a), exhibits an inverse relationship with the peak width map in Figure~\ref{resultant_maps}(c), where regions of high defect density correspond to lower intensity values and higher peak width values. This observation is quantified by the negative correlation (-0.39) between the intensity and width maps in Figure~\ref{correlation}(b), indicating that areas with significant structural imperfections exhibit broader peaks and thus lower peak intensities. 

In Figure~\ref{resultant_maps}(b), a linear gradient was observed in the peak position map, which was removed by subtraction to obtain an accurate representation of local variations in the Bragg angles. It became apparent that regions with higher Bragg angles correspond to areas of relative crystal imperfection and potential dislocations, while lower Bragg angles indicate greater crystal perfection. This relationship is further supported by the positive correlation (0.24) between the peak position and peak width maps in Figure~\ref{correlation}(c), showing that regions with higher Bragg angles tend to have broader intensity distributions, consistent with smaller crystallite sizes. However, the scientific reason for this positive correlation remains unclear. Additionally, the near-zero correlation (0.004) between the intensity and peak position maps in Figure~\ref{correlation}(a) implies no significant linear correlation.

\section{Conclusion}

In this study, we used rocking curve imaging to investigate the dislocation structures in 4H-SiC. By analyzing the integrated intensity \(I\) , peak position $\mu$, and width \(\sigma\) maps, we identified various dislocation types and lattice quality of sample. The correlation analysis revealed inverse relationship between the intensity \(I\) and width \(\sigma\) maps, indicating that regions with higher structural imperfections exhibit lower peak intensities. Additionally, a positive correlation between the peak position \(\mu\) and \(\sigma\) maps suggests that areas with higher Bragg angles tend to have broader intensity distributions, consistent with smaller crystallite sizes. These findings provide quantitative insights into the local lattice distortions and overall crystal quality, demonstrating the effectiveness of rocking curve imaging combined with Gaussian fitting in characterizing the dislocation and structural integrity of 4H-SiC crystals.

\section{Acknowledgement}

We acknowledge DESY (Hamburg, Germany), a member of the Helmholtz Association HGF, for the provision of experimental facilities. Parts of this research were carried out at PETRA III, and we would like to thank the staff for their assistance in using the P06 beamline. Beamtime was allocated for proposal ID I-20231255. This research was supported in part through the Maxwell computational resources operated at DESY.

\bibliographystyle{vancouver}
\bibliography{main}

\end{document}